\documentclass[submission]{eptcs}
\providecommand{\event}{WWV 2015} 
\usepackage{breakurl}             
\usepackage[T1]{fontenc}

\usepackage{algorithm}
\usepackage{algorithmicx}
\usepackage{algpseudocode}
\usepackage{stfloats}
\usepackage{graphicx}
\newtheorem{example}{Example} 
\newcommand{\myCall}[1]{\textnormal{\textsc{#1}}}

\title{Semantics-based Automated Web Testing}
\author{Hai-Feng Guo \qquad Qing Ouyang \qquad Harvey Siy
\institute{Department of Computer Science\\ University of Nebraska at Omaha}
\email{\{haifengguo, qingouyang, hsiy\}@unomaha.edu}
}
\def\titlerunning{Semantics-based Automated Web Testing}
\def\authorrunning{H.-F. Guo, Q. Ouyang \& H. Siy}
\begin{document}
\maketitle

\begin{abstract}
We present {\em TAO}, a
software testing tool performing automated test and oracle generation
based on a semantic approach. {\em TAO} entangles 
grammar-based test generation with automated 
semantics evaluation using a denotational semantics
framework. We show how {\em TAO} can be incorporated with 
the Selenium automation tool for automated web testing,
and how {\em TAO} can be further extended to support 
automated delta debugging, where a failing web test script can 
be systematically reduced based on grammar-directed strategies.
A real-life parking website is adopted throughout the paper
to demonstrate the effectivity of our semantics-based web testing
approach.
\end{abstract}

\section{Introduction}

As the explosive growth of web applications in the last two decades,
the demand of their quality assurance, such as requirement of reliability, 
usability, and security, has grown significantly. Software testing 
has been an effective approach to ensuring the quality of web 
applications~\cite{Sampath:2007,Lucca:2006,Stout:2001}. In practice, 
programmers and test engineers typically construct test cases either 
manually or using industrial testing automation tools such as Selenium~\cite{Selenium}, 
Watir~\cite{Watir}, and Sahi~\cite{SahiPro}. These tools provide functionalities 
for recording and replying a sequence of the GUI events as an executable 
unit test script, and provide web drivers for visualizing
testing results. However, even with the availability of these tools, 
web application testing remains difficult and time-consuming 
due to the following two observations.  
(1) Constructing web-based test scripts are mainly manual; therefore, 
obtaining sufficient test scripts with reasonable coverage to expose application failure is a challenging job. 
(2) Those practical web testing tools allow users to construct unit 
tests or test suites; however, when a failing test script, which 
can expose failure of the web application under test once executed, 
is generated, debugging and locating precise fault-inducing GUI actions
remains a tedious manual activity, since 
no further effective functionality such as automated reduction of failing 
test case is available for the purpose of automated debugging.

In this paper, we firstly introduce a declarative tool, named {\em TAO}, 
which performs automated test and oracle generation 
based on the methodology of denotational semantics~\cite{Scott:1970,Milne:1976,Schmidt:1986}. 
{\em TAO} combines our previous work on 
a grammar-based test generator~\cite{Guo:2013spe} 
and a semantics-based approach for test oracle 
generation~\cite{GuoQSIC:2014}, using a formal framework 
supporting the denotational semantics.
{\em TAO} takes as inputs a context-free grammar (CFG) and 
its semantic valuation functions, and produces test cases 
along with their expected behaviors in a fully automatic way.

Secondly, we present a new automated web testing framework 
by integrating {\em TAO} with Selenium-based web testing for 
functional testing of web applications. Our framework
incorporates grammar-based testing and semantics-based oracle generation 
into the Selenium web testing automation to generate an 
executable JUnit test suite. 
Selenium~\cite{Selenium} is an open source, robust set of tools that supports rapid development 
of test automation for Web-based applications. 
The JUnit test scripts can then be run against
modern web browsers.

Thirdly, we show {\em TAO} can be easily extended 
to support automated delta debugging.
{\em TAO} utilizes a grammar-based test generator to derive 
a structured test case based on a given CFG. We show grammatical 
structures are also valuable to reducing failing
test cases, yet maintaining syntax validity. 
Inspired by previous delta debugging approaches~\cite{Zeller:2002,Misherghi2006}, 
we present multiple grammar-directed 
reduction strategies which 
can be applied to reduce a failing web test script automatically 
based on its hierarchical structure.
On the other hand, semantics-based test oracles obtain 
expected semantic results from a recursive denotation on
each grammar structure. As a test case is reduced for 
debugging, its expected testing behaviors can also 
be adjusted simultaneously as an {\em instant oracle},
which is critical to promote automated web testing in practice.  

To demonstrate the effectiveness of our semantics-based web testing approach, 
we show in details how 
it can be applied on 
testing a real-life parking calculator website at Gerald Ford 
International Airport.

The rest of the paper is organized as follows. Section~\ref{sec:tao}
introduces our previous work on {\em TAO}.
Section~\ref{sec:taoweb} presents our web testing framework incorporating
{\em TAO} with Selenium-based web testing, and illustrates the approach
with a practical web testing example. 
Section~\ref{sec:gdd} presents a new grammar-directed 
delta debugging approach (GDD) utilizing grammar-directed reduction strategies and semantics-based instant oracles.
Section~\ref{sec:experiment} shows experimental results on web testing and automated debugging.  
Section~\ref{sec:discussion} addresses other related research work.
Finally, conclusions are given in Section~\ref{sec:conclusion}.

\section{TAO}
\label{sec:tao}

{\em TAO} is an integrated tool performing automated test and oracle generation
based on the methodology of denotational semantics.
It extends a grammar-based test generator~\cite{Guo:2013spe} 
with a formal framework supporting the three components
of denotational semantics, {\em syntax}, {\em semantics domains},
and the {\em valuation functions} from syntax to semantics.
It provides users a general Java interface
to define a semantic domain and its associated methods, which is 
integrated with {\em TAO} for supporting semantic evaluation. 
{\em TAO} takes as inputs a context-free grammar (CFG) and 
its semantic valuation functions, and produces test cases 
along with their expected behaviors in a fully automatic way.
An online version of {\em TAO} is available at~\cite{TAO2015}.

Denotational semantics~\cite{Scott:1970,Milne:1976,Schmidt:1986} is a formal methodology 
for defining language semantics, and has been widely used in language development and practical 
applications. Broadly speaking, for a web-based application under test (WUT) which requires grammar-based structured inputs,
the specification of the structured inputs is a formal language;
for those testing scripts (or methods) running together with a WUT, the
specification of those scripts is a formal language.
Denotational semantics is concerned with finding mathematical objects called domains 
that capture the meaning of an input sentence --- the expected result of the WUT, or
the semantics of a testing script --- the running behavior of the script itself along with 
the WUT. 



\begin{example}
\label{ex:java}
Consider a Java application, which takes an infix 
arithmetic expression and performs its integer evaluation. 
We use {\em TAO} to generate test inputs (arithmetic expressions)
and their expected results. 
\end{example}

To support denotational semantics, 
{\em TAO} provides a general interface 
for users to define a semantic domain and 
its associated operations as a Java class, named {\em Domain.java}. 
For Example~\ref{ex:java}, the prototype of semantic domain,
as shown in Figure~\ref{fig:ecfg}(a),  may contain
an integer variable, which will eventually hold the semantic result,
and a set of methods, such as {\em intAdd}, {\em intSub}, {\em intMul},
and {\em intDiv} supporting the basic integer arithmetic operations. 

Figure~\ref{fig:ecfg}(b) shows the input file for {\em TAO},
which contains both CFG rules and their associated semantic valuation functions. As shown in Figure~\ref{fig:ecfg}(b), 
each CFG production rule is equipped with 
a $Lisp$-like list notation denoting a semantic valuation function, 
separated by a delimiter `{\em @@}' from the CFG rule. 
A semantic valuation function, named a {\em semantic term} 
in this context, 
can be either {\em a singleton}, such as a variable in the associated 
CFG production rule or any constant,
or {\em a fully parenthesized prefix list notation} denoting an application of a valuation function, 
where the leftmost item in the list (or nested sublist) is a semantic method defined by {\em Domain.java}.

\begin{figure}[ht]
\begin{tabular}{cc}
\begin{minipage}{3in}
\begin{quote}
{\tt Semantic Domain}: {\em int}\\
{\tt Semantic Operations}: \\
    \verb+   + {\tt intAdd}: $int \times int \rightarrow int$\\
    \verb+   + {\tt intSub}: $int \times int \rightarrow int$\\
    \verb+   + {\tt intMul}: $int \times int \rightarrow int$\\
    \verb+   + {\tt intDiv}: $int \times int \rightarrow int$
\end{quote}
\end{minipage}
&
\begin{minipage}{4in}
\begin{verbatim}
 (1)      E ::= F @@ F
 (2)      E ::= E + F @@ (intAdd E F)
 (3)      E ::= E - F @@ (intSub E F)
 (4)      F ::= T @@ T
 (5)      F ::= F * T @@ (intMul F T)
 (6)      F ::= F / T @@ (intDiv F T)
 (7)      T ::= [N] @@ [N]
 (8)      T ::= (E) @@ E
 (9)    [N] ::= 1 .. 1000
\end{verbatim}
\end{minipage}
\\
(a) & (b)
\end{tabular}
\caption{(a) Semantic Domains; (b) CFG and their Valuation Functions}
\label{fig:ecfg}
\end{figure}

Consider the rule in line $(2)$; it means that 
if a test case contains a grammar structure $E+F$, 
its corresponding semantic value is denoted by a 
$\lambda$-expression $\lambda E.\lambda F.(intAdd~E~F)$, 
where the formal arguments $E$ and $F$ are omitted 
due to their implication in the CFG rule itself.
If the semantic term is a singleton (e.g., in line $(1)$), 
it simply returns the semantic result of the singleton; 
otherwise, it triggers an associated operation (e.g., $intAdd(E, F)$)  
as defined in the domain class, assuming the semantic values of $E$ and $F$
have been obtained recursively. Note that the occurrences of $E$ and $F$ on the right 
of `{\em @@}' denote their respective semantic values,
and the variable {\tt [N]} is a symbolic terminal, denoted by a pair of squared brackets, 
representing a finite domain of integers, from $1$ to $1000$.

\subsection{Tagging Variables}

In automated test script generation, it would be ideal that runtime assertions can be 
automatically embedded into a test script, so that when a test script is invoked for 
software testing, the running result immediately indicates either success or failure 
of testing; otherwise, a post-processing procedure is typically required to check the running result against the oracle. 

{\em TAO} provides an easy tagging mechanism for 
users to embed expected semantic results 
into a generated test case. It allows users to create a tagging variable 
as a communication channel for passing results from semantics generation to test generation.
A tagging variable is in a form of $\$[N]$, where $N$ can be any non-negative
integer. A tagging variable can be defined in front of any semantic term 
{\tt <SemTerm>}, either a singleton
or a fully parenthesized prefix list notation, in a form of $\$[N]${\tt : <SemTerm>}. 

\begin{example} 
\label{ex:td}
If we add the following two grammar rules into the beginning of the CFG 
in Figure~\ref{fig:ecfg}, 

\begin{verbatim}
         TD ::= E Assert @@ $[1] : E
     Assert ::= '=' $[1]  
\end{verbatim}

\noindent
where {\tt TD} is the new main CFG variable deriving 
an arithmetic expression and its expected evaluation result as well.
Thus, we may get a sample test case: $3*(8-4)=12$, where $12$ is the expected
semantic value obtained by the tagging variable, $\$[1]$.
\end{example}

Each tagging variable has its application scope on deriving a test case. 
The rule (1) for {\tt TD} allows the tagging variable $\$[1]$ to record the value 
of the semantic term $E$, and allows any occurrences of $\$[1]$ to be replaced by
its recorded value within the scope of deriving ``{\tt E Assert}'' during test generation.
{\em TAO} allows users to define 
multiple tagging variables in a single semantic function 
for both catching semantic results and
embedding runtime assertions. Intermediate semantic values can also be recorded and embedded into test scripts.

\section{TAO-based Web Testing Framework}
\label{sec:taoweb}

\begin{figure}[ht]
	\centering
		\includegraphics[width=0.82\textwidth]{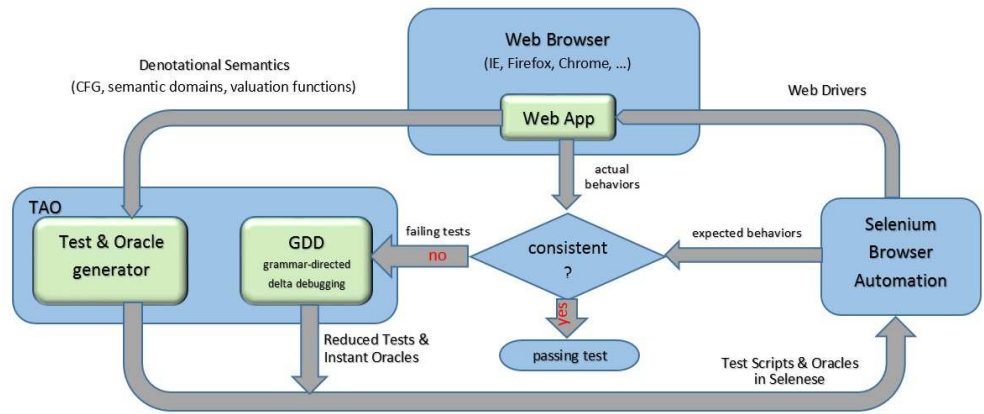}
	\caption{Automated Web Testing Framework}
	\label{fig:webframework}
\end{figure}

Figure~\ref{fig:webframework} presents an automated web testing framework 
based on our testing tool {\em TAO} and Selenium browser automation. The framework
consists of the following main procedures.
(i) A WUT is modeled using a methodology of 
denotational semantics, where CFGs are used to represent the GUI-based execution 
model of the WUT, semantics domains are used to describe functional behaviors 
of the WUT, and valuation functions map user interactions to expected web behaviors. 
(ii) {\em TAO} takes the denotational semantics of the WUT as an input,
and automatically generates a suite of JUnit tests, 
supported in the Selenium browser automation tool.
Each JUnit test contains a GUI scenario of the WUT as well as expected 
WUT behaviors embedded.
(iii) Through Selenium's web drivers, a suite of JUnit test scripts 
can be executed to test different scenarios of the WUT. The actual running 
behaviors of the WUT will be automatically collected to compare against its 
expected behaviors for consistency checking. 
(iv) Once a failing test script is found, that is, when running a test script,
its actual behaviors are inconsistent from its expected ones, {\em TAO} will invoke
a grammar-directed delta debugging strategy to repeatedly reduce the failing 
test case to a {\em minimized} one for automated debugging.
In this section, we will address the first three procedures in details, and
the procedure (iv) will be explained in the following Section~\ref{sec:gdd}.  

Selenium~\cite{Selenium} is an open source, robust set of tools that 
supports rapid development of test automation for Web-based applications. 
It provides a test domain-specific language, named {\em Selenese}, 
to write test scripts in a number of popular programming languages, 
including Java. The test scripts can then be run against most modern 
web browsers through Selenium's web drivers.

\subsection{A Web Application under Test --- A Parking Calculator }

\begin{figure}[ht]
\centering
\begin{tabular}{cc}
\includegraphics[width=0.55\textwidth]{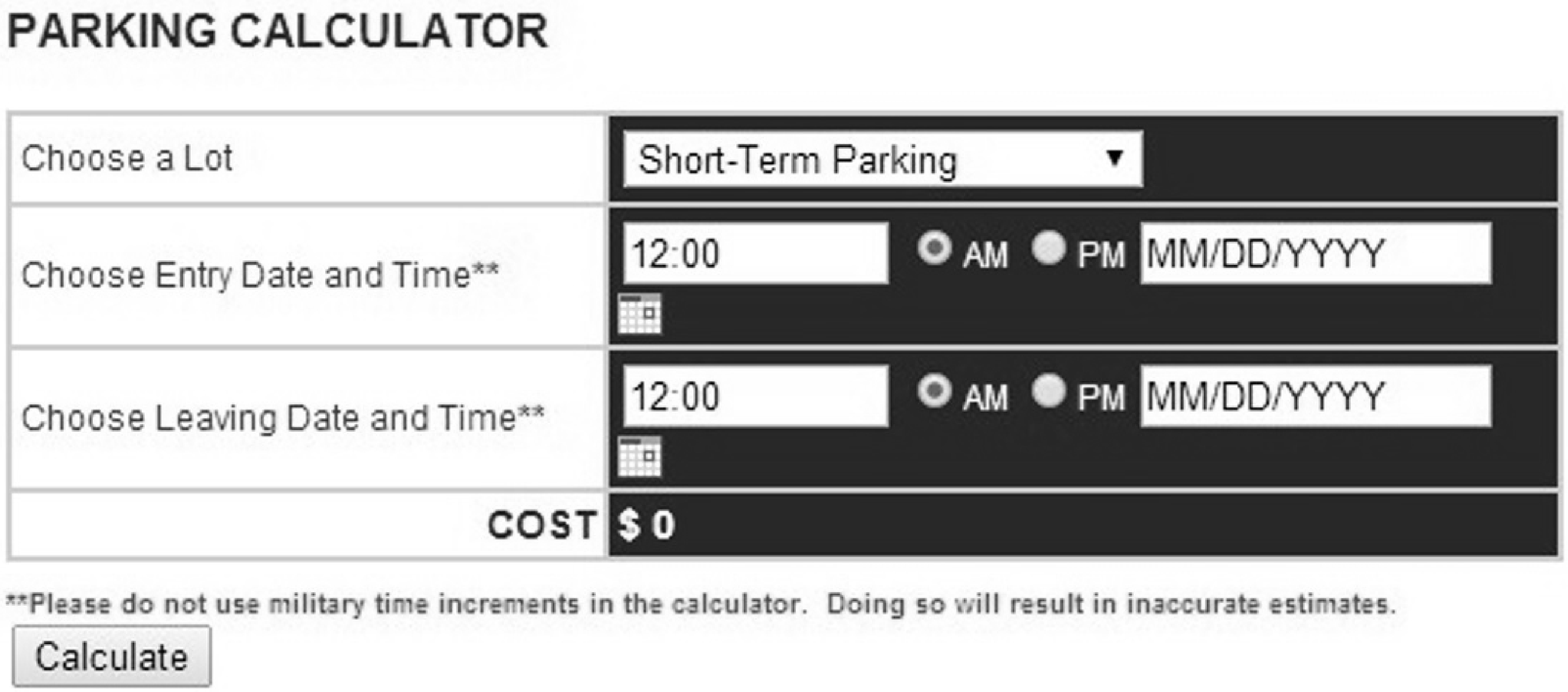}& {\hspace{.1in}\includegraphics[width=0.44\textwidth]{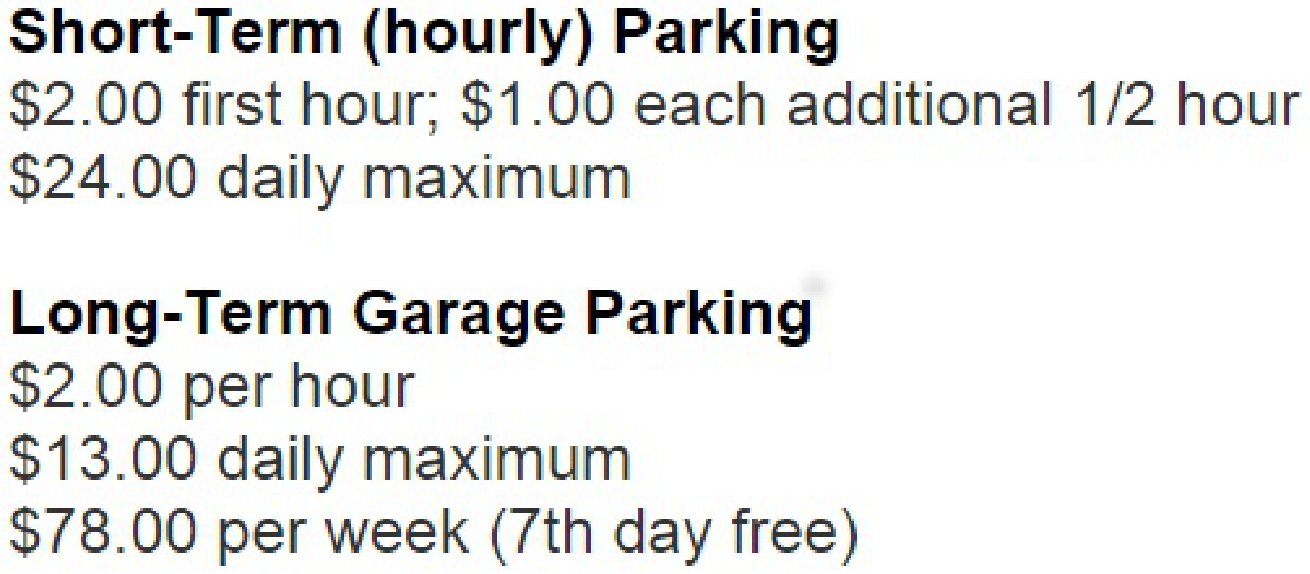}} \\
(a) Parking Calculator & (b) Partial 2014 Parking Rates
\end{tabular}
\caption{Parking Calculator and Rates}
\label{fig:parking}
\end{figure}

We use a parking calculator website at Gerald Ford International 
Airport\footnote{\tt Parking Calculator Web: www.grr.org/ParkCalc.php; \\
\indent\indent Parking Rates Web: www.grr.org/ParkingRates.php} 
to demonstrate how our automated web testing framework works
practically, integrating the automation of {\em TAO} with Selenium.
Figure~\ref{fig:parking}(a) shows the parking calculator GUI 
on the website, where users can select a parking lot type, 
entry and exit dates and times, and press the ``Calculate'' button.

Figure~\ref{fig:parking}(b) shows a part of parking rates,
for the {\tt short-term} and {\tt long-term garage} parking lots,
adopted by the Gerald Ford International 
Airport in $2014$.

\subsection{WUT Execution Model in Denotational Semantics}

We show how to follow the methodology of denotational semantics to 
specify the user-web interactions of the WUT execution model 
and their expected behaviors. To catch the semantics of web operations in the parking calculator, we 
define the following semantic domains and necessary operations in {\em Domain.java},

\begin{quote}
{\tt Semantic Domains}: \\
    \verb+   + {\bf Price} = $double$ \\
    \verb+   + {\bf Duration}, {\bf DTsf} = $long$ \\
    \verb+   + {\bf Time}, {\bf Date}, {\bf LotType} = $string$ \\
    \verb+   + {\bf AmPm} = $boolean$ \\
    \verb+   + {\bf Hour}, {\bf Minute}, {\bf Month}, {\bf Day}, {\bf Year} = $int$ \\
{\tt Semantic Operations}: \\
    \verb+   + {\tt price}: {\bf LotType} $\times$ {\bf Duration} $\rightarrow$ {\bf Price}\\
    \verb+   + {\tt sfSub}: {\bf DTsf} $\times$ {\bf DTsf} $\rightarrow$ {\bf Duration}\\
    \verb+   + {\tt simpleFmt}: {\bf Time} $\times$ {\bf Date} $\rightarrow$ {\bf DTsf}\\
    \verb+   + {\tt date}: {\bf Month} $\times$ {\bf Day} $\times$ {\bf Year} $\rightarrow$ {\bf Date}\\		
    \verb+   + {\tt time}: {\bf Hour} $\times$ {\bf Minute} $\rightarrow$ {\bf Time}\\
    \verb+   + {\tt time24Fmt}: {\bf AmPm} $\times$ {\bf Time} $\rightarrow$ {\bf Time}
\end{quote}

\noindent
where the semantic operation {\tt time} takes inputs 
\verb+<+$hour$\verb+>+ and \verb+<+$minute$\verb+>+ 
and returns a {\bf Time} string in the form of 
``\verb+<+$hour$\verb+>+/\verb+<+$minute$\verb+>+/00''; 
{\tt time24Fmt}
transform the time string into a 24-hour format by considering
the {\tt am} or {\tt pm}  option;
{\tt date} returns a {\bf Date} string in the format of 
``\verb+<+$month$\verb+>+/\verb+<+$day$\verb+>+/\verb+<+$year$\verb+>+'';
{\tt simpleFmt} combines a {\bf Date} string and a 24-hour 
{\bf Time} string into a {\bf DTsf}/$long$ type
using the Java {\em SimpleDateFormat} package;
{\tt sfSub} calculates the duration in a long type 
from entry to exit in {\em SimpleDateFormat};
and {\tt price} calculates the total parking fee based 
on the lot type and the duration.

As shown in Figure~\ref{fig:parking}(a), 
the typical web operation sequence is to 
(i) choose a parking lot
type: {\em short-term}, {\em economy}, {\em surface}, 
{\em valet}, or {\em long-term} garage; 
(ii) choose entry date and time; (iii) choose leaving date and time, 
and (iv) press the ``Calculate'' button.
Such a sequence of user-web interactive operations can be described
using a CFG as partially shown in Figure~\ref{fig:parkingCFG}, where each grammar rule
is followed by a semantic valuation function, separated by ``@@''. 

\begin{figure}[ht]
\hspace{.5in}
\begin{minipage}{3in}
\footnotesize
\begin{verbatim}
Operations ::= Lot Duration Cal @@ (price Lot Duration)
Lot ::= Short | Economy | Surface | Valet | Garage
Short ::= 'new Select(driver.findElement(By.id("Lot")))
          .selectByVisibleText("Short-Term Parking");' @@ short
Duration ::= Entry Exit @@ (sfSub Exit Entry)
Entry ::= EnTime EnDate @@ (simpleFmt EnTime EnDate)
Exit ::= ExTime ExDate @@ (simpleFmt ExTime ExDate)
EnTime ::= AmPm EnTimeInput @@ (time24Fmt AmPm EnTimeInput)
EnDate ::= 'driver.findElement(By.id("EntryDate")).clear;
     driver.findElement(By.id("EntryDate")).sendKeys("' TDate '");' @@ TDate
TDate ::= [Month] '/' [Day] '/' [Year] @@ (date [Month] [Day] [Year])
[Month] ::= 1..12
EnTimeInput ::= 'driver.findElement(By.id("EntryTime")).clear();                   
     driver.findElement(By.id("EntryTime")).sendKeys("' TTime '");' @@ TTime
...  ...
Cal ::= 'driver.findElement(By.name("Submit")).click();'
\end{verbatim}
\end{minipage}
\caption{Partial CFG and Valuation Functions}
\label{fig:parkingCFG}
\end{figure}

Each CFG rule is followed by a valuation function which evaluates the expected
semantics based on its syntactic structure by calling pre-defined semantic 
operations in {\em Domain.java}, such that when a test script, a sequence
of Selenese statements, is derived in TAO, its expected behavior on the parking
calculator is automatically evaluated as a corresponding test oracle.
Note that for a unit CFG rule without semantic valuation functions defined, 
for example, the rule ``{\tt Lot ::= Short}'', it relays the semantic value
of {\tt Short} to its parent rule, that is, equivalent to ``{\tt Lot ::= Short @@ Short}''.

Given the CFG and associated semantic valuation functions in Fig~\ref{fig:parkingCFG},
{\em TAO} is expected to generate a suite of JUnit test scripts, each of which consists of 
a sequence of Selenese statements to simulate a scenario of
users' operations on the parking calculator website. 
Thus, a terminal in the CFG 
should be a legal Selenese statement, which utilizes a Selenium web driver to
communicate with web browsers. 
For example, consider the CFG rule for {\tt Cal} in Fig.~\ref{fig:parkingCFG}.
It actually simulates a user's operation clicking the {\em Calculate} button. 

For conciseness, we only show the CFG for the typical web operation sequence;
in practice, we also consider the possible permutation among operations. For example,

{\small
\begin{verbatim}
      Operations ::= Lot Duration Cal @@ (price Lot Duration)
      Operations ::= Duration Lot Cal @@ (price Lot Duration)
      Duration ::= Entry Exit @@ (sfSub Exit Entry)
      Duration ::= Exit Entry @@ (sfSub Exit Entry)
\end{verbatim}
}

Furthermore, we are able to generate each test case with one or more rounds of 
continuous parking cost calculations, each of which is followed by a 
runtime check as shown below

{\small
\begin{verbatim}
      Test ::= Round
      Test ::= Round Test
      Round ::= Operations Fetch Assert @@ $[1] : Operations
      Fetch ::= 'actualResult = driver.findElement(By.cssSelector("b")).getText();'
      Assert ::= 'if (!consistent(actualResult, $[1])) fail();'
\end{verbatim}
}


\noindent
where the variable {\tt Test} denotes a complete test case, which
can possibly contain multiple rounds, each denoted by the variable {\tt Round},
of parking calculations. Additionally, the variable {\tt Operations} is used to 
specify a sequence of basic operations calculating a round of parking cost, 
{\tt Fetch} is used to specify a Selenese Java statement
fetching the actual parking cost from the web at runtime, 
and {\tt Assert} is derived to a Java statement, by calling
a pre-defined Java method {\tt consistent}, 
to compare the fetched actual cost with
the expected cost generated by TAO, and reveal the testing failure
if the costs are inconsistent. Note that the tagging variable \$[$1$]
is used to hold the semantic value of {\tt Operations}, the expected parking cost,
and embedded into the CFG definition of {\tt Assert} as a part of test script.  

To generate an executable Java JUnit test script through Selenium's web drivers, 
each test script was then combined with a standard Selenium JUnit test
header and footer to form a complete JUnit test script. 

\section{Grammar-directed Delta Debugging}
\label{sec:gdd}

When a JUnit test script fails a runtime consistency check, for example,
when actual costs calculated by the airport online parking calculator is different
from the expected test oracle, we call such a test script a {\em failure-inducing} 
test case. In this section, we show how {\em TAO} can be extended to support
automated delta debugging to reduce failure-inducing test cases. 
{\em TAO} utilizes a grammar-based test generator
to derive a structured test case. Grammatical structures
are also valuable to reducing failure-inducing test cases to better 
understand the software failure. Test case reduction based on syntax 
is critical to make it sure that reduced test cases are syntactically valid;
as a test case is reduced, its expected semantics or oracle on software testing 
will be changed as well, since denotational semantics typically maps a syntactic
structure into mathematical domains. 
In this section, we show how our semantics-based test oracle approach
advance automated delta debugging.

\subsection{Delta Debugging}
\label{sec:dd}

Delta debugging (DD)~\cite{Zeller:2002} has been a popular 
automated debugging approach to simplifying and isolating failure-inducing 
inputs for fault localization.
It simplifies a failing test case to a {\em minimal} one that still produces
the testing failure, where the minimization is defined in terms that any
further desired simplification of the test case would make the testing 
succeed. {\em DD} assumes 
a set of {\em changeable circumstances} and uses a general binary search
within those changeable circumstances.   
However, identifying changeable circumstances in a test case often requires syntactic 
information so that any involved change will not invalidate the test case itself.
Additionally, even if a set of changeable circumstances have been successfully identified, 
their heterogeneity may prevent us 
applying a simple binary search. 


Contrast to {\em DD}, hierarchical delta debugging (HDD)~\cite{Misherghi2006}
parses a test case into a hierarchical structure of changeable circumstances
based on its syntactic information so that the {\em DD} technique can be applied on 
each structural level to maintain syntactic validity.
However, such a hierarchical structure adopted in HDD is 
not a traditionally defined parse tree, 
but a reorganized structure suitable for the application of {\em DD}.
Constructing such a hierarchical structure may need a domain-specific parser. 

\subsection{Grammar-directed Test Reduction}

{\em TAO} utilizes a given CFG to derive structured test case. 
Such a CFG is also valuable to reducing 
failure-inducing test cases for automated debugging,
not only for the purpose of syntactic validity, but
also for providing clues how the reduction can be
automated in a systematic way.
We use the following example to illustrate grammar-directed
test reduction strategies.

\begin{example}
\label{ex:program}
Consider the following partial CFG for a simple structural programming language.
Each program contains  a definition part
(denoted by {\tt Def}) and a sequence of statements (denoted by {\tt StmtSeq}), 
such as assignment, if, or loop statements. 
{
\begin{verbatim}
                Program  ::= Def StmtSeq
                StmtSeq* ::= Stmt
                StmtSeq  ::= Stmt StmtSeq
                   Stmt  ::= while Cond { StmtSeq }
                   ... ...
\end{verbatim}
}
\end{example}

Now we present multiple grammar-directed reduction strategies as follows, 
which can be applied to reduce a test case 
while still following syntactic validity.
\smallskip

\noindent
{\bf [Reduction by Default]}: {\em TAO} allows users optionally to 
specify a default grammar rule by simply marking an asterisk (*) 
after the defining variable in a rule. 
For example, ``\verb+StmtSeq* ::= Stmt+'' is a default rule,
which typically means that {\tt Stmt} is one of the simplest yet valid structures 
for {\tt StmtSeq}. The reduction strategy by default searches 
for each node labeled by {\tt StmtSeq} in the derivation tree,
and checks whether its child nodes can be simplified based on 
the default rule. 

\begin{figure}[ht]
	\centering
  \includegraphics[width=0.80\textwidth,height=.9in]{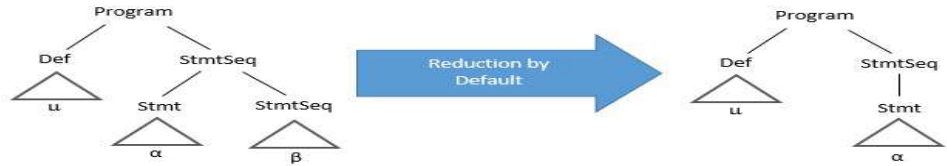}
	\caption{Reduction by Default}
	\label{fig:default}
\end{figure}

Assume that a failing test program $\mu\alpha\beta$ is found 
whose corresponding derivation tree is shown in Figure~\ref{fig:default}, 
where $\mu$, $\alpha$, and $\beta$, 
respectively, represent subtrees for a definition part, a single statement, and 
a sequence of statements. 
With the reduction strategy by default, the reduced derivation tree would still
be a valid one syntactically, corresponding to its reduced program 
$\mu\alpha$ and the derivation as follows:

{
\[
Program \Rightarrow Def~StmtSeq \Rightarrow^{*} \mu~StmtSeq \Rightarrow
\mu~Stmt \Rightarrow^{*} \mu\alpha
\]
}


\vspace{-.15in}
\noindent
{\bf [Reduction by Direct Recursion]}: 
Similarly considering ``\verb+StmtSeq ::= Stmt StmtSeq+'', 
a directly recursive rule, {\em TAO} can search
for each occurrence of the {\tt StmtSeq} node, and check
whether its child nodes involve a recursive node. 

\begin{figure}[ht]
	\centering
  \includegraphics[width=0.80\textwidth,height=.9in]{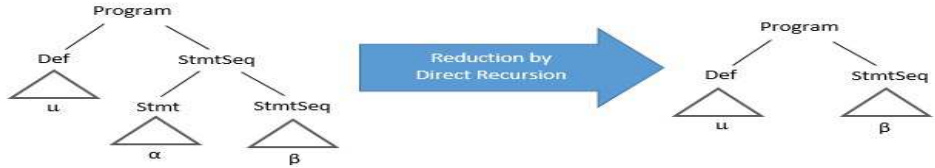}
	\caption{Reduction by Direct Recursion}
	\label{fig:directRecursion}
\end{figure}

\noindent
If so, as shown in Figure~\ref{fig:directRecursion}, 
{\em TAO} can reduce the original derivation into 
a reduced one, corresponding to a reduced test 
program $\mu\beta$ and its valid derivation as follows:
{
\[
Program \Rightarrow Def~StmtSeq \Rightarrow^{*} \mu~StmtSeq \Rightarrow
\mu~Stmt \Rightarrow^{*} \mu\beta
\]
}

\vspace{-.15in}
\noindent
{\bf [Reduction by Indirect Recursion]}: We often see that some CFG variables 
are defined in an indirectly recursive way. 
Consider the partial CFG in Example~\ref{ex:program}.
The variable {\tt StmtSeq} contains an indirectly recursive 
definition through:
{
\[
StmtSeq \Rightarrow Stmt \Rightarrow while~Cond~\{~StmtSeq~\} 
\]
}

\begin{figure}[ht]
	\centering
  \includegraphics[width=0.80\textwidth,height=1in]{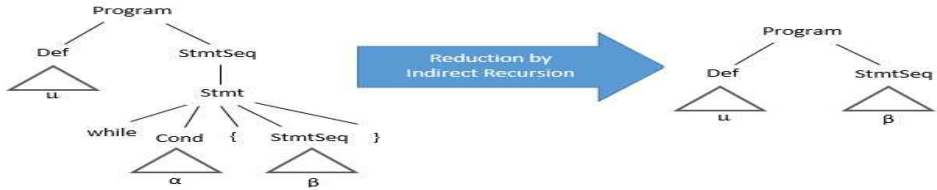}
	\caption{Reduction by Indirect Recursion}
	\label{fig:IDR}
\end{figure}

\vspace{-.2in}
\noindent
Thus, we may have the reduction strategy by indirect recursion 
as shown in Figure~\ref{fig:IDR}. It
reduces the derivation into a valid one in a similar
way as the reduction strategy by direct recursion, but 
searches for alternative reduction of {\tt StmtSeq} 
in a much deeper way within its derivation subtree.  

All three reduction strategies we have just introduced are based on the assumption
that a failure-inducing reduced test case probably gives a 
better intuition to locate the faults or 
understand the failure causes in software testing.  
In practice, the reduction strategy by indirect recursion 
may compromise runtime efficiency 
because it is unknown that which variable actually has an indirect
recursion and looking for indirect recursion over the whole derivation 
tree is expensive. Therefore, {\em TAO} provides users a declarative way 
to specify a list of applicable reduction strategies. 
Consider Example~\ref{ex:java} for reducing 
failure-inducing arithmetic expressions; users may specify a list of reduction
strategies as follows:

{
\begin{verbatim}
      TAO-reduction: {"default", "directRec", "indirectRec: {E,F,T}"}
\end{verbatim}
}

\noindent
For the sake of efficiency, users need to explicitly list those 
CFG variables which are both defined using indirect recursion
and used for reduction purposes. 

\subsection{Semantics-based Instant Oracle}

For automated delta debugging, grammar-directed reduction 
helps to maintain the syntactic validity on reduced test cases.
As a failure-inducing test case is reduced, its expected semantics 
or oracle on software testing needs to be instantly updated as well 
so that further automated reduction can be continuously performed 
to {\em minimize} failure-inducing patterns 
for precise fault localization. 

\begin{figure}[ht]
	\centering
		\includegraphics[width=0.6\textwidth]{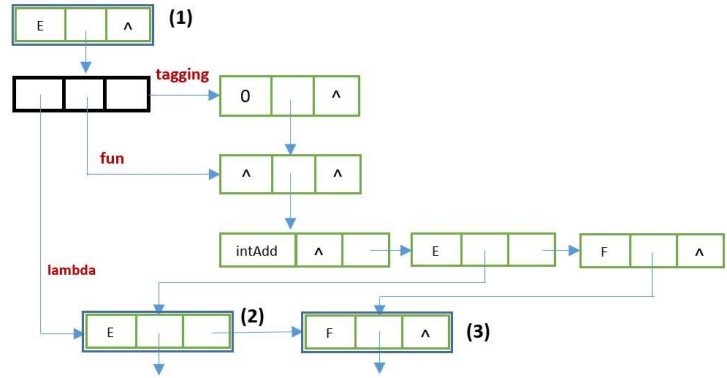}
	\caption{A semantic valuation function for {\tt E ::= E + F}}
	\label{fig:intAdd}
\end{figure}

{\em TAO} has been extended with an instant oracle mechanism for 
dynamic test case reduction. Each test case generated by {\em TAO} 
comes with a derivation tree, which can be further manipulated by applying 
any of the reduction strategies, if applicable. 
A semantic tree is dynamically built up for obtaining an instant oracle
by applying pre-defined valuation functions (e.g., as shown
in Figure~\ref{fig:ecfg}) on a reduced derivation tree.
To support instant oracles, {\em TAO} stores semantic valuation functions
into a mapping set indexed by each CFG rule. Figure~\ref{fig:intAdd}
shows an example of data structures for storing semantics valuation functions,
corresponding to a partial derivation:
{
\[E^{(1)} \Rightarrow E^{(2)} + F^{(3)}.\]
}

\vspace{-.15in}
\noindent
where a variable with a superscript (e.g., $E^{(1)}$) tells that the variable 
is bound with a semantic node (also highlighted by a double-line node) 
with the same label as shown in Figure~\ref{fig:intAdd}. 
The tagging node $0$, a feature in {\em TAO}
but irrelevant in this paper, defines a default
tagging variable \$[$0$] catching the semantic value. 

\subsection{Grammar-directed Delta Debugging (GDD)}

We present a new delta debugging algorithm, {\em GDD}, which 
incorporates grammar-directed reduction strategies 
with the instant oracle generator in a search based procedure. 
As shown in Algorithm~\ref{alg:GDD},
{\em GDD} takes a failure-inducing test case, {\em test}, as an input, 
and repeatedly applies each applicable reduction strategy to obtain
a reduced one (lines $6-8$) until no further reduction 
is possible (lines $9-12$) in the recursion.
The sub-function \myCall{GetDerivationTree}($current$) (line $5$) 
returns the root of the derivation tree associated with $current$;   
the sub-function {\tt APPLY($\alpha$, $current$)} (line $7$) applies the reduction
strategy $\alpha$ on the derivation tree of $current$ in a search-based way,
and returns a reduced one if applicable, otherwise returns the same $current$ 
test case.  

\begin{algorithm}[ht]
\footnotesize
\begin{algorithmic}[1]
\State {\bf Input:} {\em test}, a failure-inducing test case
\State {\bf Output:} a {\em reduced} failure-inducing test case
\Function{GDD}{$test$}
   \State $current \gets test$
	 \State $root \gets$ \Call{getDerivationTree}{$current$}
	 \For{each applicable reduction strategy $\alpha$}
	     \State $current \gets$ \Call{apply}{$\alpha$, $current$, $root$, $root$}
	 \EndFor
 	 \If{($current$ is different from $test$)}
	     \State \Return \Call{gdd}{$current$}
	 \Else
	     \State \Return $current$
	 \EndIf
\EndFunction
\end{algorithmic}
\caption{The GDD approach}
\label{alg:GDD}
\end{algorithm}

\begin{algorithm}[!ht]
\footnotesize
\begin{algorithmic}[1]
\makeatletter\setcounter{ALG@line}{14}\makeatother
\State {\bf Input:} (1) $\alpha$, a grammar-directed reduction strategy; (2) {\em test}, a failure-inducing test case;
\State ~~~~~~~~~~~~(3) $root$, the root of the derivation tree of {\em test}; (4) {\em pNode}, a node in the derivation tree
\State {\bf Output:} a {\em reduced} failure-inducing test case
\Function{APPLY}{$\alpha$, $test$, $root$, $pNode$}
   \If{($pNode$ is a CFG terminal)} 
		  \State \Return $test$  
	 \Else \Comment{$pNode$ is a CFG variable}
	    \If{($\alpha$ is applicable on $pNode$)} 
	       \State {\em store} the first child link of $pNode$ \Comment{use first-child/next-sibling data structure}
				 \State \Call{reduceBy}{$pNode$, $\alpha$}
				 \State $reduced \gets$ \Call{getTestCase}{$root$}
				 \State $oracle \gets$ \Call{InstantOracle}{$root$}
				 \If{(\Call{testing}{{\em SUT}, {\em reduced}, {\em oracle}} fails)}
				    \State $test \gets reduced$  \Comment{still failure-inducing}
				 \Else
				    \State {\em restore} the first child link of $pNode$
				 \EndIf
			\EndIf
	 \EndIf 
	 \For{each child node $cNode$ of $pNode$}
	     \State $test \gets$ \Call{apply}{$\alpha$, $test$, $root$, $cNode$}
	 \EndFor
   \State \Return $test$
\EndFunction
\end{algorithmic}
\caption{{\bf APPLY}: a search-based reduction procedure}
\label{alg:apply}
\end{algorithm}

The function  \myCall{APPLY}($\alpha$, $test$, $root$, $pNode$), defined 
in Algorithm~\ref{alg:apply}, applies the reduction strategy $\alpha$ 
on each node $pNode$ in the derivation tree rooted at $root$ in a 
top-down, depth-first order.
For each non-terminal node $pNode$ (line $21$), 
the sub-function checks whether the the reduction 
strategy $\alpha$ is applicable on the subtree rooted 
at $pNode$ (line $22$).
We highlight two implementation details in this algorithm.
(1) We adopt the first-child/next-sibling data structure for 
representing derivation trees; thus, reducing 
a subtree of $pNode$ can be achieved by changing its first
child link.
(2) We use a store/restore mechanism to maintain the original
subtree of $pNode$ (line $23$). In case that the reduced test case is 
not failure-inducing, we have to restore the original subtree
of $pNode$ (lines $27-30$); otherwise, 
the reduced one will be used for further reduction. 
The function \myCall{reduceBy}($pNode$, $\alpha$) 
applies the reduction strategy $\alpha$ on $pNode$;
\myCall{getTestCase} and \myCall{instantOracle} 
return a test case and its oracle corresponding
to the derivation tree rooted at $root$, respectively.
The function \myCall{testing}($SUT$, $reduced$, $oracle$)
is invoked to check whether the $reduced$ test case
is still failure-inducing. Only a failure-inducing $reduced$ test
case will be kept for further delta debugging.
In both cases, either reduced or not, the
function will continue with applying the reduction strategy
$\alpha$ on each child node of $pNode$ recursively (lines $34-36$). 

\begin{example}
\label{ex:expReduce}
Assume that the Java application under test, as described in Example~\ref{ex:java},
handles arithmetic in a right-associative way instead of a left-associative by mistake,
but it respects the precedences of operators. For example, given a test case 
``$2 * (5 - 3 + 4)$'', the Java application returns a wrong result $-4$ 
instead of $12$, due to the wrong handling order of ``$5-3+4$''.  Now we illustrate how 
our GDD is able to reduce the failure-inducing test case, given the CFG and valuation 
functions shown in Figure~\ref{fig:ecfg}, where the CFG rules (1)(4)(7) are default
rules of the variables {\tt E}, {\tt F}, and {\tt T}, respectively.
\end{example}

\begin{figure}[ht]
	\begin{tabular}{cccc}
		\includegraphics[width=0.25\textwidth]{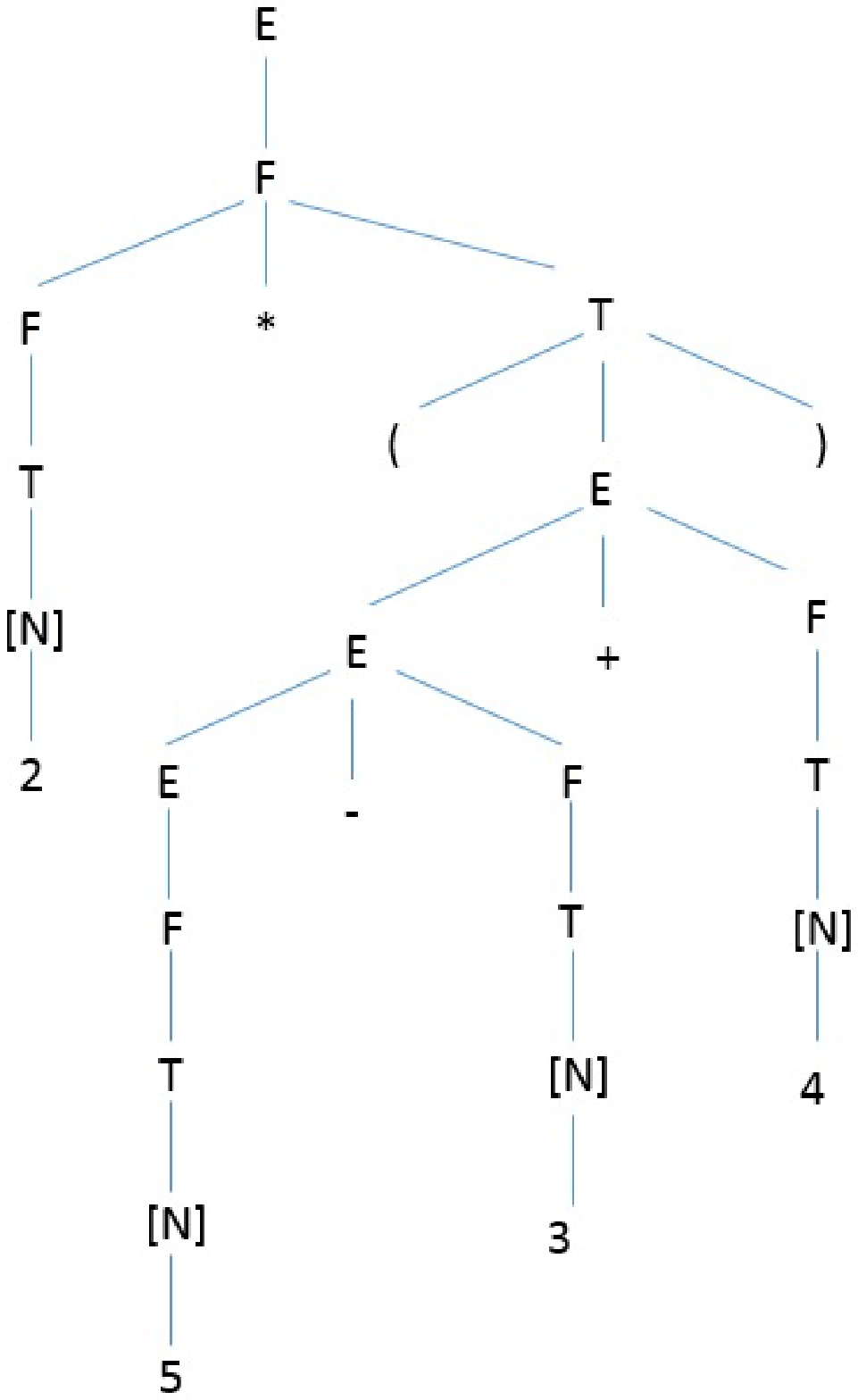} &
		\includegraphics[width=0.25\textwidth]{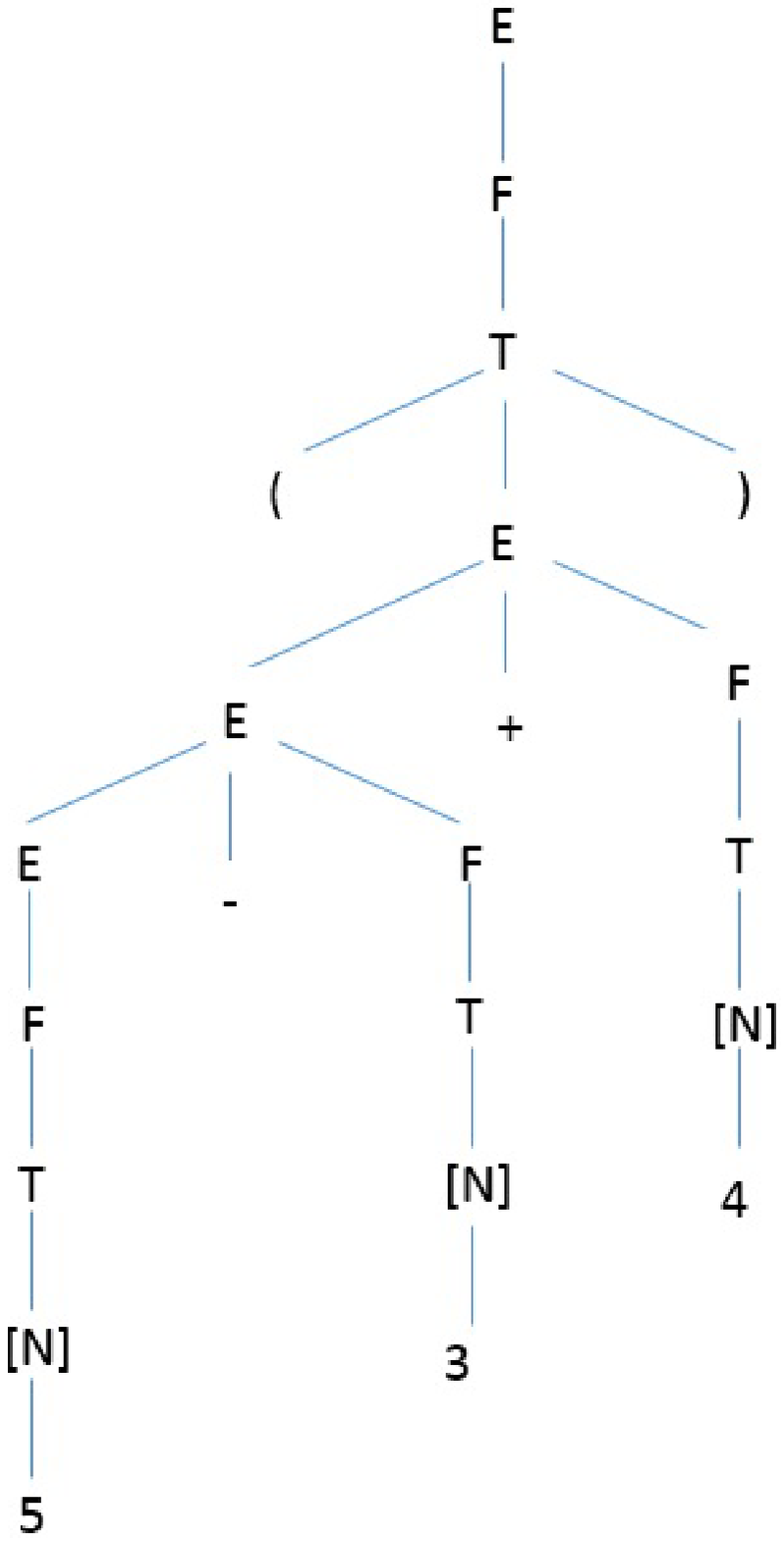} &
		\includegraphics[width=0.17\textwidth]{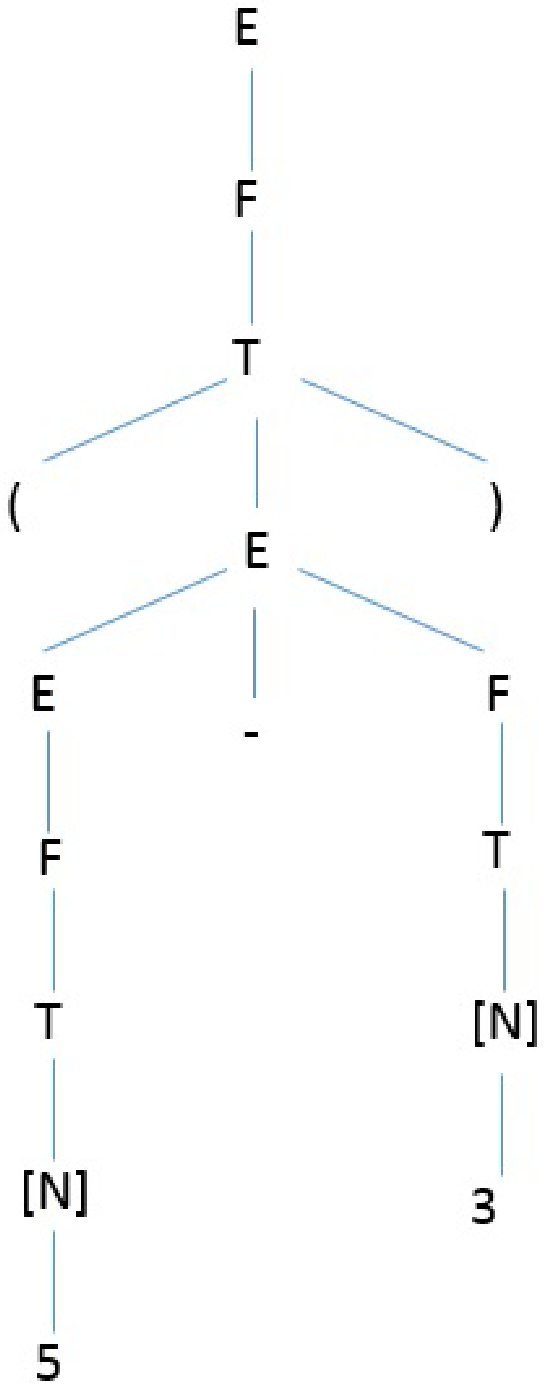} &
		\includegraphics[width=0.25\textwidth]{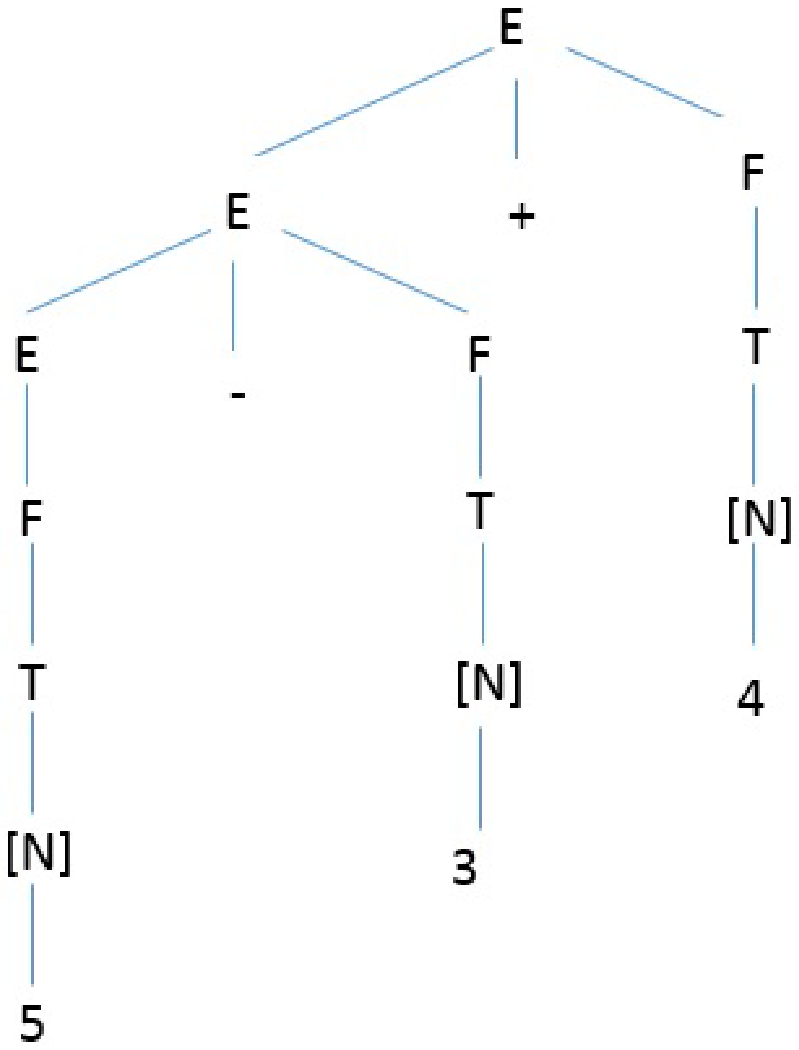} \\
   (a) $2*(5-3+4)$ & (b) $(5-3+4)$ & (c) $(5-3)$ & (d) $5-3+4$ 
    \end{tabular}
	\caption{An Example of Reducing Failure-inducing Test Cases}
\label{fig:expReduce}
\end{figure}

Figure~\ref{fig:expReduce}(a) shows the derivation tree for the original failure-inducing
test case $2 * (5 - 3 + 4)$, following the given CFG in Figure~\ref{fig:ecfg}.
By applying the reduction strategy by default, the {\tt APPLY} function is 
able to reduce the derivation tree in Figure~\ref{fig:expReduce}(a) to a simplified yet
still failure-inducing one as shown in Figure~\ref{fig:expReduce}(b), 
due to the fact that {\tt F ::= T} is a default rule for {\tt F}.  
Further reduction by default rules from the derivation tree in Figure~\ref{fig:expReduce}(b)
is possible (e.g., some occurrences of {\tt E} can be derived to {\tt F} directly based
on the default rule of {\tt E}), however, none of these reduced one by default rules
is failure-inducing. 

Similarly, from the derivation tree in Figure~\ref{fig:expReduce}(b), no further
reduced failure-inducing test case is able to be generated by applying the reduction
strategy by direct recursion. For example, the derivation tree in Figure~\ref{fig:expReduce}(c) is a reduced one applying the reduction strategy by direct recursion of {\tt E};
however, the reduced test case $(5-3)$ is not a failure-inducing one, unable to expose
the fault of left-associativity.

By further applying the reduction strategy by indirect recursion of {\tt E}, 
the {\tt APPLY} function is able to reduce the derivation tree 
in Figure~\ref{fig:expReduce}(b) to a simplified one in 
Figure~\ref{fig:expReduce}(d), which corresponds to a maximally-reduced yet 
precise failure-inducing pattern $5-3+4$. 
The three grammar-directed reduction strategies
can be applied repeatedly until no more further reduction can be made. 

\section{Experimental Results}
\label{sec:experiment}

In this section, we first show our preliminary experimental
results on automated delta debugging by applying the GDD approach
on applications which require structured inputs, 
and then show experimental results on Selenium-based web testing.

\subsection{Locating failure-inducing patterns on buggy Java programs}

Preliminary experiments have been conducted on testing and 
debugging $5$ different buggy Java programs 
(student submissions) which take an arithmetic 
expression as an input and perform its integer calculation, as 
described in Example~\ref{ex:java}.
We used the extended {\em TAO} 
with new capabilities, instant oracle and 
grammar-directed reduction strategies, 
to generate $1000$ arithmetic 
expressions and locate the failure-inducing patterns.
{\em TAO} takes the following inputs: a file of 
CFGs associated with its semantic functions~\footnote{The input file 
is shown in Fig.~\ref{fig:ecfg}, with the addition that CFG rules (1)(4)(7) 
are marked with asterisks to
denote that they are default rules for variables {\tt E}, {\tt F}, and {\tt T}, 
respectively.} and a list of reduction strategies to be applicable, 

{
\begin{verbatim}
   TAO-reduction: {"default", "directRec", "indirectRec: {E,F,T}"}.
\end{verbatim}
}

Table~\ref{tbl:grading} shows our experimental results
of reduced failure-inducing patterns by applying our {\em GDD} approach. 
Consider the first Java program. For example,
``$88+(45*15+(15/85*99/88*27/95-92+22)*96*13/67/48)$'' is a generated 
failure-inducing test case; 
that is, given this expression input, the actual evaluation result returned 
by the Java program is different from the expected result 
in the oracle generated by {\em TAO}.
Our GDD approach is able to reduce
the failure-inducing input to ``$15/85/88$'', which 
implies the simplified failure-inducing pattern $//$, as shown in the 
Table~\ref{tbl:grading}. By collecting all the simplified
failure-inducing patterns, we are able to speculate that
the first Java program may not handle right-associativity properly. 
\begin{table}[ht]
  \centering
 \caption{Failure-inducing Patterns and their Causes via GDD} 
 \begin{tabular}{c|c}
  \hline
    programs & failure-inducing patterns ({\em Possible Causes})  \\
    \hline
		1 &  $\{\verb#-+, /*, */, //, --#\}$~~({\em right-associativity}) \\
    \hline
    2 & $\{\verb#()#\}$~~({\em parenthesis not properly handled}) \\
    \hline
    3 & $\{\verb#/-, /+, /*, */, +/, *-, -/, *+, //, --#\}$ \\
		  & ({\em\small right-associativity and operator precedence})\\
    \hline
    4 & $\{\verb#-*-, -/-, -/+, -*+#\}$~({\em partial operator precedence ignorance}) \\
		\hline
    5 & $\{\verb#/-, *-, *+, /+#\}$~~({\em operator precedence ignorance}) \\
    \hline
       \end{tabular}
        \label{tbl:grading}
\end{table}

The following table shows average reduction ratios on 
the lengths of failure-inducing inputs by applying the GDD approach on debugging the $5$ buggy Java programs. 
For example, for the second program, buggy due to parentheses 
issues, the lengths of failure-inducing test inputs can be reduced by $87\%$ on average.
The overall average reduction ratio among $5$ programs 
is by $81\%$ on lengths of failure-inducing inputs. Automated instant oracle generation plays a key role in automating 
debugging, specifically in identifying and
reducing failure-inducing inputs.
{\begin{table}[ht]
\centering
 \caption{Average Reduction Ratio on failure-inducing Inputs} 
 \begin{tabular}{c|c|c|c|c|c|c}
  \hline
	programs & 1 & 2 & 3 & 4 & 5 & average \\
	\hline
	reduction ratio & $80\%$ & $87\%$ & $80\%$ & $79\%$ & $79\%$ & $81\%$ \\
  \hline 
 \end{tabular}
\label{tbl:ratio}
\end{table}
}

\subsection{Selenium-based Web Test Script Reduction}

The second experiment shows how the extended {\em TAO}
can be used for Selenium-based web testing by incorporating 
semantics-based testing into Selenium web testing framework to 
generate an executable {\em Selenese} JUnit test suite and  using 
GDD for automated debugging.
We use the parking calculator website at Gerald Ford International 
Airport for the experiment, where the CFGs and valuation functions 
are partially shown in Figure~\ref{fig:parkingCFG}. 
The experiment utilizes {\em TAO} 
to generate web test scripts and their associated oracles, compares actual 
web testing results with expected oracles, 
and reveals testing failure automatically. 
We collected a suite of $500$ JUnit web scripts generated by {\em TAO}. 
Our  experimental results reveals that the average failure ratio is about $11.24\%$.

We further applied our GDD approach on reducing failing test JUnit scripts, giving 
a list of reduction strategies specified as follows:
\begin{center}
\verb+TAO-reduction: {"default", "directRec"}.+
\end{center}

We further used $200$ executable Selenium-based test scripts for the experiment of 
automated web testing and debugging by applying the GDD approach, where
$28$ executable test scripts cause testing failure. Each of those failing
test scripts may contain one or multiple rounds of parking cost calculations,
and in each round of parking cost calculation, users may set entry/exit dates and times
in any order and modify them repeatedly.  

Our GDD approach was able to reduce a failing test script to 
a simplified one, with an average reduction ratio about $22\%$. 
We found out that most failures were caused by different time-boundary issues. 
For example, consider the short-term parking rates, where 
the daily maximum short-term parking fee is $\$24$; however,
the web parking calculator could display $\$26$ if your total 
parking time is $12$ hours and $30$ minutes. We summarize the
faults as follows:

\begin{table}[ht]
  \centering
 \caption{Faults Summary for the Online Parking Calculator} 
 \begin{tabular}{c|l}
 \hline
 Lot Types & \hspace{1in}Faults \\
 \hline
 Garage,  & 1. weekly maximum was violated \\
 Surface,                              & 2. daily maximum was violated \\
	Economy														& 3. wrong parking cost was given when the leave time is\\
	& ~~~earlier than the entry time \\
 \hline
  Short-term & 4. daily maximum was violated \\
	           & 5. half hour price was not properly calculated \\
	\hline
	Valet  & 6. wrong parking cost was given when the leave time is\\
	& ~~~earlier than the entry time \\
	\hline
 \end{tabular}
 \end{table}

\smallskip
Both automated instant oracle generation and grammar-directed delta debugging 
are critical to automating web testing and fault localization. 

\section{Other Related Works and Discussions}
\label{sec:discussion}

\noindent
[{\bf Grammar-based Test Generation }]  
Grammar-based test generation (GBTG) provides a systematic approach to
producing test cases from a given context-free grammar.
Unfortunately, naive GBTG is
problematic due to the fact that exhaustive random test case production 
is often explosive. Prior work on GBTG mainly relies
on explicit annotational controls, such as 
production seeds~\cite{Sirer:1999}, combinatorial control parameters~\cite{Lammel:2006}, and extra-grammatical annotations~\cite{Hoffman:2011}.
However, GBTG with explicit annotational controls is not only 
a burden on users, but also causes unbalanced testing coverage, 
often failing to generate many corner cases. 

{\em TAO} takes a CFG as input, requires zero annotational control from users, 
and produces well-distributed test cases in a systematic way.
{\em TAO} guarantees (1) the termination of test case generation, 
as long as a {\em proper} CFG, which has no inaccessible variables
and unproductive variables, is given; and (2) 
that every generated test case is
{\em structurally different} as long as the 
given CFG is unambiguous.

\noindent
[{\bf Model-based Web Testing}] 
Many previous researches on automated testing of web application use model-based web testing, such as using finite state machines~\cite{Andrews:2005}, a model of application state space~\cite{Mesbash:2009},
or an application's event space~\cite{Saxena:2010}, to name a few. 
These approaches rely on a heuristic approach for generating
test cases based on an application model, but generally
seek extra assistance to maintain good testing coverage. For example, 
the Artemis tool~\cite{Artzi:2011} incorporates a model-based testing with 
a feedback-directed strategy for automated testing of web applications.
However, these testing approaches for web applications, mainly focusing
on test generation automation, lack a further mechanism for automating
test oracle generation and delta debugging.  

Our semantics-based automated web testing also belongs to the category of
model-based web testing, since grammar-based test generation typically 
uses CFGs to describe a structured input data model or a user-web interactive
behavior model.

\noindent
[{\bf Automated Delta Debugging}]   
Artzi, et al.~\cite{artzi10} proposed a white-box testing technique, which
monitors the execution of the WUT to record symbolic path constraints, 
and then uses model checking to generate test inputs for dynamic web applications. 
The resulting tool, {\em Apollo}, can further {\em minimize} the set of constraints
which lead to the failure-inducing inputs by intersecting sets of constraints among
failing-inducing inputs. Our GDD approach is a black-box testing technique
for general users, who may not have the source code of the WUT, but are able to 
generate test scripts for testing web applications. The GDD approach 
can be used to reduce either test inputs or test scripts in a grammar-directed 
systematic way. {\em TAO} combines grammar-based test generation, semantics-base oracle 
generation, and grammar-directed delta debugging as an integrated tool.

\section{Conclusions}
\label{sec:conclusion}

We presented TAO, a
testing tool performing automated test and oracle generation
based on a semantics-based approach, 
and showed a new automated web testing framework by 
integrating {\em TAO} with Selenium-based web testing for 
web testing automation. Our framework is able to generate
a suite of executable JUnit test scripts by utilizing 
grammar-based test generation and semantics-based oracle 
generation. 

The semantics-based web testing approach is also valuable to promote 
automated delta debugging, as it provides sufficient flexibility
on supporting grammar-directed reduction strategies and 
semantics-based instant oracle generation. We extend {\em TAO}
with a new grammar-directed delta debugging approach (GDD) for automated
delta debugging. As shown in experiments, not only can {\em TAO} 
be used to reduce and locate failure-inducing input patterns
for those applications which require structured inputs,
but it can also reduce web test scripts 
to assist fault localization.

\section{Acknowledgements}

We want to thank the anonymous referees for their valuable comments 
that improved the presentation. This research project has been partially supported 
by the First Data Corporation.

\bibliographystyle{eptcs}
\bibliography{main}
\end{document}